\documentclass[10pt,pra,aps,twocolumn,showpacs]{revtex4}
\usepackage[dvips]{graphicx}
\usepackage{amsmath,amssymb,setspace}

\begin{document}

\title{Adiabatic Floquet Picture for Hydrogen Atom in an Intense Laser Field}
\author{Yujun Wang}
\altaffiliation{Present address: JILA, University of Colorado, 440 UCB, Boulder, Colorado, 80309, USA}
\author{J.~V.~Hern\'andez}
\author{B.~D.~Esry}
\affiliation{J.R. Macdonald Laboratory, Kansas State University, Manhattan,
Kansas, 66506}

\begin{abstract}
We develop an adiabatic Floquet picture in the length gauge to describe the
dynamics of a hydrogen
atom in an intense laser field. In this picture, we discuss the roles played by
frequency and intensity in terms of adiabatic potentials and
the couplings between them, which gives a physical and intuitive picture for
quantum systems exposed to a laser field.
For simplicity, analyze hydrogen and give the adiabatic potential curves as well
as some physical quantities that can be
readily calculated for the ground state.  Both linearly and circularly polarized
laser fields are discussed.
\end{abstract}
\maketitle

\section{\bf Introduction}

The dynamics of atoms in intense, ultrashort laser pulses involves the
absorption and emission of many photons, making non-perturbative methods
necessary.
By far the most common such method is the direct numerical solution of the
time-dependent Schr\"odinger equation.
Unfortunately, this approach does not often provide the kind of simple physical
picture that lets one understand the dynamics and predict what will
happen under different circumstances.  Moreover, numerically solving the
time-dependent Schr\"odinger equation accurately is, in most cases, a demanding
task.
These shortcomings have played a large role in promoting the three-step (or simpleman's) model~\cite{Corkum1993,KSK1992}
of ionization dynamics.  In its normal application, each step of the this simple model
draws from a different theoretical formulation, although it can be rigorously derived with a
few approximations~\cite{Lewenstein1994}.

At the same time, the non-perturbative treatment of molecular dissociation in
intense, ultrashort laser pulses has led to a different --- but simple and useful -- physical picture that
guides many researchers' understanding of these systems.  This picture combines
the Floquet representation with the Born-Oppenheimer approximation, producing
Born-Oppenheimer potentials that include the effects of the laser field.  All of
the insight earned in understanding field-free molecular dynamics can thus be
directly applied to molecules in a laser field.  Importantly, these ``dressed
potentials'' provide a description of the laser-induced dynamics using only a
single theoretical formulation.  These dressed potentials have also provided the
interpretations behind the mechanisms of bond-softening, vibrational trapping
(bond-hardening), above-threshold dissociation, and zero photon dissociation.
Recently, the picture has even been generalized to include
ionization-dissociation channels, leading to the prediction of a new mechanism
--- above threshold Coulomb explosion --- which was subsequently
measured~\cite{Esry,AStaud:PRL2007,BEsry:PRA2010}.

A natural question to ask, then, is whether such a Floquet approach can be
applied to atoms in intense fields and, if so, does it yield a similarly useful
physical picture~\cite{JHern:BAP2003}. In this paper, we answer this question by generating dressed
potentials for the electron's motion in a hydrogen atom exposed to an intense
laser.
It is the generation of dressed potentials that distinguishes our work from the substantial
body of work using the Floquet approach for atoms in intense fields~\cite{Chu_review}.
Our approach does, however, bear a closer resemblance to treatments of Rydberg
atoms~\cite{JShir:PR1965,RJens:PR1991,ABuch:PR2002,HMaed:PRL2006}
and is very similar in spirit to the recent work of Miyagi and Someda~\cite{HMiyagi2010}.
They emphasized the comparison of circular and linear polarization in the velocity
and acceleration gauges.  While they use the adiabatic potentials for radial
motion in the acceleration gauge to interpret their numerical results, we
focus on developing the length gauge adiabatic potentials into a more general framework
for understanding and predicting the response of an atom to an intense laser pulse.
We further test the quantitative predictive power of the potentials by calculating physical observables
in the adiabatic approximation.

\section{\bf Theoretical Background}
\label{Theory}

The Floquet theorem has been invoked many times in the past to solve the
time-dependent Schr\"odinger equation for an atom in an intense field. In fact,
Floquet calculations provide some of the most accurate non-perturbative
ionization rates available.  With one primary exception, though, no simple
picture like that for molecules has emerged from this previous work. Floquet
ideas did, however, provide an elegant understanding for that exception: the
stabilization of atoms against ionization in a high-frequency laser field~\cite{Intro1}.
We will thus be applying the Floquet theorem in a non-standard way.

For simplicity, and to allow the focus to be on the new representation we
introduce, we will consider a CW laser field.  This simplification has the
additional benefit of allowing easy quantitative comparison with previous
calculations.  Our approach can be generalized to the case of a laser pulse
using the ideas described in~\cite{Chu_review}.

The Hamiltonian for a hydrogen atom in an intense laser field is given in atomic
units by
\begin{equation}
H=-\frac{1}{2}\nabla^2-\frac{1}{r}+\boldsymbol{r}\cdot \boldsymbol{\cal E},
\label{Hamil}
\end{equation}
using the dipole approximation and choosing the length gauge (see
App.~\ref{App_Gauges}).
For a linearly polarized CW laser field, the electric field is $\boldsymbol{{\cal
E}}={\cal E}_0\cos(\omega t)\hat{z}$.
In the above expressions, $r$ and $z$ are the electron's coordinates with the
$z$ axis chosen along the laser polarization;
${\cal E}_0$ is the amplitude of the laser's electric field; and $\omega$ is the
laser frequency.

Since the Hamiltonian is periodic, $H(t+2\pi/\omega)=H(t)$, the Floquet theorem
states that the solutions of the time-dependent Schr\"odinger equation
\begin{equation}
i\frac{\partial}{\partial t}\psi(\boldsymbol{r},t)=H\psi(\boldsymbol{r},t)
\label{Schrodinger}
\end{equation}
have the form
\begin{equation}
\psi(\boldsymbol{r},t)=e^{-i\varepsilon t}\phi(\boldsymbol{r},t),
\end{equation}
where $\phi(\boldsymbol{r},t)=\phi(\boldsymbol{r},t+2\pi/\omega)$.
Upon substitution into Eq.~(\ref{Schrodinger}), one finds the eigenvalue
equation
\begin{equation}
\left [ H-i\frac{\partial}{\partial t}\right
]\phi(\boldsymbol{r},t)=\varepsilon\phi(\boldsymbol{r},t)
\label{Eff_Schrodinger}
\end{equation}
This equation shows that, in a sense, the Floquet approach allows us to treat
time like a coordinate.  For later convenience, we define the effective
Hamiltonian ${\cal H}$ as
\begin{equation}
{\cal H}=H-i\frac{\partial}{\partial t}.
\label{Hamil_Eff}
\end{equation}

The primary reason that the molecular Floquet approach described above is so
useful is that it produces easy-to-interpret potentials that include the effects
of the laser field. So, our goal is to similarly produce dressed potentials for
an atom.  The most natural way to do this is to take $r$ to be an adiabatic
parameter and solve
\begin{equation}
{\cal H}_{\rm ad}\Phi_{\nu}=U_{\nu}(r)\Phi_{\nu}.
\label{Ad_Eq}
\end{equation}
In this case, the adiabatic effective Hamiltonian is
\begin{equation}
{\cal H}_{\rm ad}=\frac{L^2}{2 r^2}-\frac{1}{r}+\boldsymbol{r}\cdot
\boldsymbol{\cal E}-i\frac{\partial}{\partial t}
\label{Hamil_Eff_Ad}
\end{equation}
where $L^2$ is the squared orbital angular momentum operator.  The eigenstates
from Eq.~(\ref{Ad_Eq})
form a complete set at every $r$ so that the wave function $\phi$ can be written
without approximation as
\begin{equation}
 \phi(\boldsymbol{r},t) = \sum_\nu \frac{1}{r} F_\nu(r)
\Phi_\nu(r;\theta,\varphi,t).
\label{AdExpansion}
\end{equation}
The eigenvalues $U_\nu(r)$ of Eq.~(\ref{Ad_Eq}) are precisely the potentials we
seek.  That they serve this purpose can be seen by substituting
Eq.~(\ref{AdExpansion}) into Eq.~(\ref{Eff_Schrodinger}) which yields the
following equations for $F_\nu(r)$
\begin{widetext}
\begin{equation}
\left[-\frac{1}{2}\frac{d^2}{dr^2}+U_\nu(r)\right] F_\nu(r)
-\frac{1}{2} \sum_{\nu'}
\left[2P_{\nu\nu'}(r)\frac{d}{dr}+Q_{\nu\nu'}(r)\right]F_{\nu'}(r) = \varepsilon
F_\nu(r).
\label{RadEqns}
\end{equation}
\end{widetext}
These equations are identical to those one obtains in the Born-Oppenheimer
representation, and the derivative coupling matrices ${\bf P}$ and ${\bf Q}$ are
defined as usual by
\begin{align}
P_{\nu\nu'}(r) &= \langle\!\langle \Phi_\nu | \frac{d}{dr} | \Phi_{\nu'}
\rangle\!\rangle \nonumber \\
Q_{\nu\nu'}(r) &= \langle\!\langle \Phi_\nu | \frac{d^2}{dr^2} | \Phi_{\nu'}
\rangle\!\rangle .
\label{NonAd}
\end{align}
The double bracket notation indicates that these matrix elements should be
integrated over one cycle of the field in time as well as over the electron's
angles.  Note that the coupled radial equations (\ref{RadEqns}), including the
potentials and coupling, are time-independent.  And, in the absence of the
field, the equations decouple, yielding the usual separable solutions for a
hydrogen atom.  It is worth noting that this adiabatic representation is, in
principle, exact.

To solve the adiabatic equation (\ref{Ad_Eq}), we expand the spatial degrees of
freedom in $\Phi$ on the spherical harmonics.
We can guarantee the time periodicity of $\Phi$ --- and thus of $\phi$ --- by
expanding its time dependence via the discrete Fourier transform. We thus have:
\begin{equation}
  \Phi_\nu(r;\theta,\varphi,t)=\sum_{nlm}C_{nlm}^\nu(r) Y_{lm}(\theta,\varphi)
  e^{-i n \omega t},
  \label{Fourier_Decomp}
\end{equation}
explicitly showing the parametric dependence on $r$.  The Floquet index $n$
ranges from $-\infty$ to $\infty$; $l$ is the angular momentum; and $m$ is the
projection of the angular momentum on the $z$ axis.  In this basis, the matrix
elements of ${\cal H}_{\rm ad}(r)$ for a linearly polarized laser field are
\begin{widetext}
\begin{equation}
\langle\!\langle nlm |{\cal H}_{\rm ad}|n'l'm'\rangle\!\rangle=\delta_{n
n^\prime}\delta_{l l^\prime}\delta_{m m^\prime}
\left[\frac{l(l+1)}{2r^2}-\frac{1}{r}-n\omega\right]+(\delta_{n,n'+1}+\delta_{n,
n'-1})\frac{1}{2}r {\cal E}_0 \langle Y_{l m}|\cos\theta|Y_{l' m'}\rangle.
\label{Adia_Hamil_LP}
\end{equation}
\end{widetext}
The term $\langle Y_{l m}|\cos{\theta}|Y_{l' m'}\rangle$ can be evaluated
analytically in terms of $3j$ symbols.
The diagonal terms $\langle\!\langle nlm |{\cal H}_{\rm ad}|nlm\rangle\!\rangle$
are the diabatic potentials.

We can simplify the potential curves by using the dipole selection rules
implicit in Eq.~(\ref{Adia_Hamil_LP}).
Taking the initial state to be the field-free hydrogen ground state and choosing
it to correlate with $n=0$ in the field, only the diabatic channels that couple
to the $(n,l,m)=(0,0,0)$ channel are relevant.
Therefore only those diabatic
channels with both $n$ and $l$ even or
$n$ and $l$ odd are allowed.
Further, all channels with non-zero $m$ can
be eliminated, so this label will be suppressed.  The physical
meaning of the Floquet index is now clear, it represents the net number of
photons absorbed or emitted by the hydrogen atom in the laser.

Before proceeding to the results and discussion,
we note that by selecting the length gauge the whole matter-field interaction is
included in ${\cal H}_{\rm ad}$ since it is local in $r$. In the velocity gauge,
however, this is not the case since the matter-field interaction contains a term
with a derivative in $r$ which is not included in ${\cal H}_{\rm ad}$. While
${\cal H}_{\rm ad}$ in the acceleration gauge does include all the terms
involving the field, it is also not ideal for reasons that are discussed in
App.~\ref{App_Gauges}. To be clear, even in the length and acceleration gauges
in which all terms involving the field are included in ${\cal H}_{\rm ad}$, it
is {\em not} true that all effects of the field are included in the adiabatic
potentials --- the non-adiabatic couplings are also field-dependent. From this
discussion, it is clear that the adiabatic potentials are gauge-dependent, as
are the non-adiabatic couplings.   However, this gauge-dependence is just a
matter of
the distribution of physical information between potentials and couplings; any
physical observable calculated in any gauge will be properly gauge-independent
if calculated exactly, {\em i.e.} by including all of the potentials and their
couplings.

\section{\label{sec:Results}Results and Discussion}
\subsection{Linearly Polarized Fields}
\subsubsection{Potential Curves}
\begin{figure}
 \includegraphics[clip=true,scale=0.6]{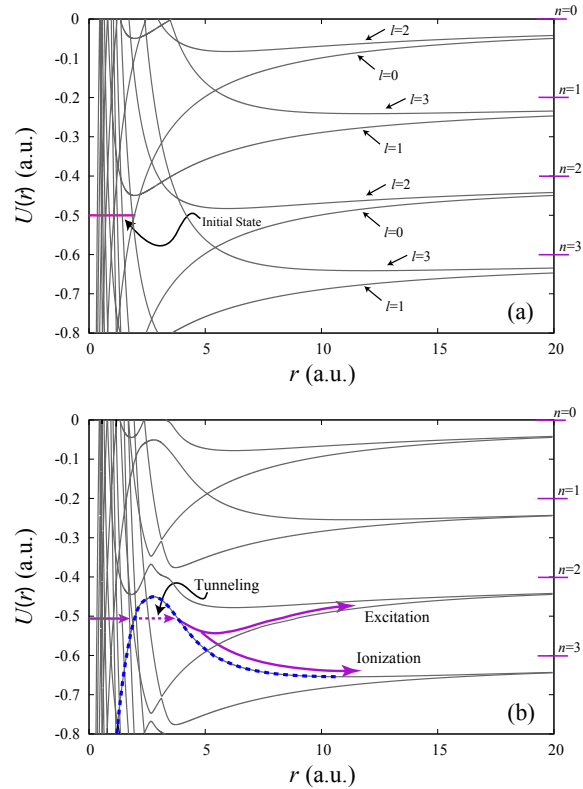}
\caption{(color online.) (a) The diabatic Floquet potentials for a hydrogen atom at $\omega=0.2$\,a.u.~
in a linearly polarized laser field.
The thick blue curve denotes the initial channel.
(b) The corresponding adiabatic Floquet potentials at, ${\cal E}_0=0.05$\,a.u.
The Floquet blocks included here are $-10\leq n\leq 10$ and the angular momentum
states included are $l=0,1,2,3$.
}
\label{fig:pot_curves}
\end{figure}
Figure~\ref{fig:pot_curves}(a) shows the diabatic potential curves, including
only the lowest two $l$s for each $n$.  The diabatic potentials are thus
partially dressed in that they have been shifted by the net number of photons
exchanged with the field. They do not, however, reflect any further effects of
the field as all such effects lie in the couplings in this representation.
Passing to the adiabatic representation by diagonalizing ${\cal H}_{\rm ad}$
produces fully dressed potentials that include much more of the effects of the
laser.  These adiabatic potentials are shown in Fig.~\ref{fig:pot_curves}(b),
and each crossing of diabatic potentials from Fig.~\ref{fig:pot_curves}(a) has
now become an avoided crossing.  A transition between diabatic channels
represents the absorption or emission of a photon or photons (according to
$\Delta n$) as does passing through an avoided crossing on the {\em same}
adiabatic channel.  Transitions between adiabatic channels, which occur
predominantly at avoided crossings, may or may not involve the exchange of
photons with the field.

Although Fig.~\ref{fig:pot_curves}(b) has more potentials than the typical
molecular example, the parallels between the atomic and molecular cases are
rather clear.  For instance, taking the initial state to be the field-free
hydrogen $1s$ state (indicated in both panels of Fig.~\ref{fig:pot_curves} at an
energy of {$-0.5$~a.u.}), we see from Fig.~\ref{fig:pot_curves}(b) that the
electron can tunnel through the barrier induced by the field to emerge on
potentials correlating to the $n$=2 and 3 thresholds.  In the process of
tunneling, the wavepacket thus absorbed three $\omega$=0.2~a.u.~photons.

If the tunneled wavepacket reaches $r\rightarrow\infty$ on one of the $n$=3
channels, then it has ionized --- the analog of bond-softening in molecules.
The final kinetic energy of such an electron is the difference between the
initial energy and the final threshold energy, which in this case is 0.1~a.u.~
Since the electron absorbed two photons to ionize, dipole selection rules
dictate that it should end up with $l$=1 or 3.
>From the diabatic potentials in Fig.~\ref{fig:pot_curves}(a), we can identify the
most likely pathway as being $(n,l)=(0,0)\longrightarrow (1,1) \longrightarrow (2,2)
\longrightarrow (3,3)$ by following the crossings.  Comparison with the adiabatic
potentials in Fig.~\ref{fig:pot_curves}(b) show that these are the channels
that form the barrier through which the electron tunnels to ionize.  We thus predict
that the ionized electron will primarily have $l=5$.
Physically, then, we expect the angular distribution of
photoelectrons to have primarily $f$-wave character.

If the wavepacket ends up on the $n$=2 channel instead, then the electron has
simply been excited since the $n$=2 threshold energy lies above the initial
energy.  To get to the $n$=2 threshold, the wavepacket must emit a photon at the
crossing around $r$=5~a.u.  Based on the small gap at that crossing compared to
the kinetic energy at that $r$, the electron will most likely traverse the
crossing diabatically and thus ionize.  Should it emit a photon and remain bound
however, arguments similar to the ionization case predict that the excitation
pathway should lead mainly to $l$=0 rather than $l=2$.

\begin{figure*}
 \includegraphics[clip=true,scale=0.4]{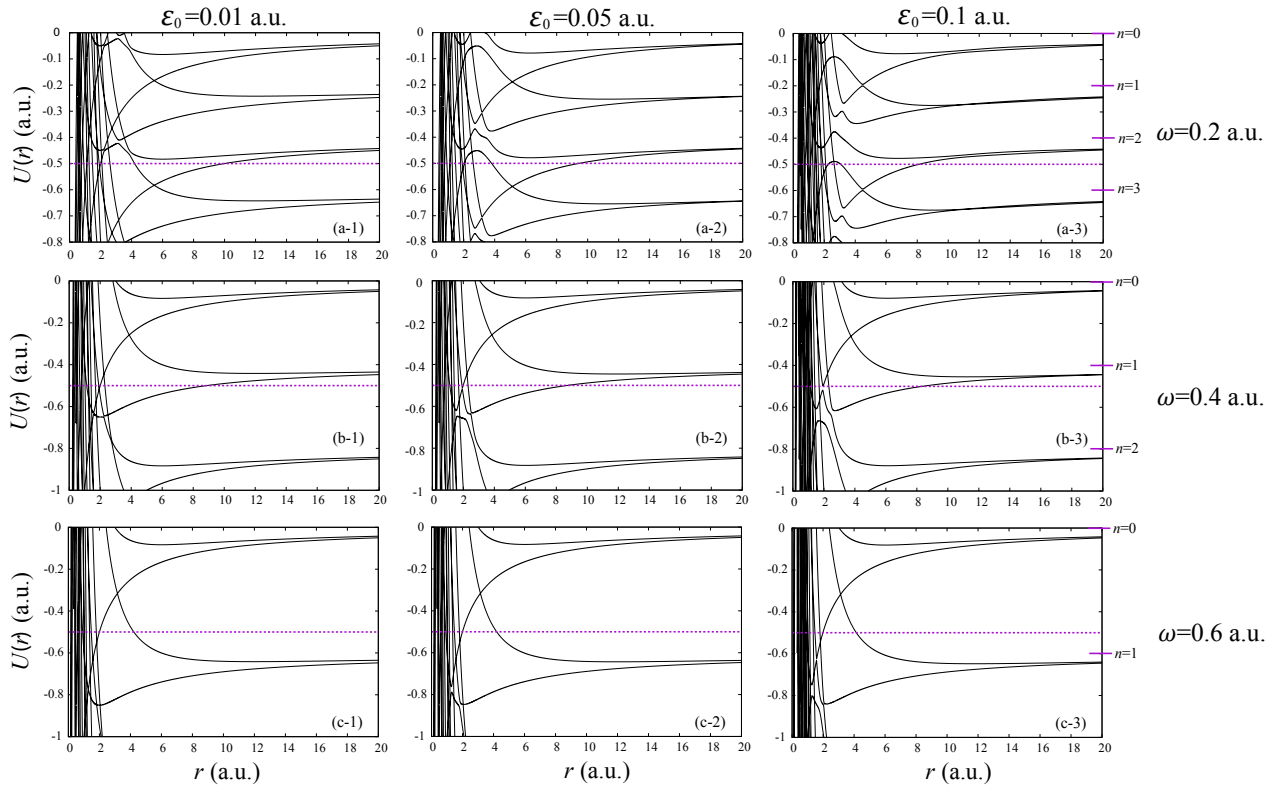}
\caption{The adiabatic Floquet potentials for a hydrogen atom in a linearly polarized laser field at
(a-1) $\omega=0.2\,$a.u., ${\cal E}_0=0.01$\,a.u.;
(a-2) $\omega=0.2$\,a.u., ${\cal E}_0=0.05$\,a.u.; (a-3) $\omega=0.2$\,a.u.,
${\cal E}_0=0.05$\,a.u.; (b-1) $\omega=0.4$\,a.u., ${\cal E}_0=0.01$\,a.u.;
(b-2) $\omega=0.4$\,a.u., ${\cal E}_0=0.05$\,a.u.; (b-3) $\omega=0.4$\,a.u.,
${\cal E}_0=0.1$\,a.u.; (c-1) $\omega=0.6$\,a.u., ${\cal E}_0=0.01$\,a.u.
(c-2) $\omega=0.6$\,a.u., ${\cal E}_0=0.05$\,a.u.; (c-3) $\omega=0.6$\,a.u.,
${\cal E}_0=0.1$\,a.u. Seven
Floquet blocks, $n=-3$ to 3, and four angular momentum states are used for these curves.}
\label{fig:pot_curves_ads}
\end{figure*}

These adiabatic potentials thus provide considerable qualitative --- and
quantitative --- information.  In fact, within this picture, we can also
distinguish simultaneous absorption or emission processes from sequential
processes.  Along the excitation pathway described above, for instance, the
process is initiated by the simultaneous absorption of three photons during the
tunneling, followed after some delay by the emission of one photon.
``Simultaneous'' in this case means that the photons were exchanged with the
field at a single avoided crossing.  The three-photon absorption during
tunneling really takes places over an $r$ range of about 1.5~a.u. as the
potential has clear $n$=0 character only up to about $r$=2~a.u., and $n$=3
character only after about $r$=3.5~a.u.  The electron, of course, takes a finite
time to cross this distance.  This example is rather extreme since most avoided
crossings are much more localized as can be seen in Figs.~\ref{fig:pot_curves}
and \ref{fig:pot_curves_ads}.

Being able to identify where the transitions occur in $r$ could be useful for
the attosecond control of the electron's dynamics.  The travel time of the
electron between crossings can be estimated, for instance, making it possible to
identify opportunities for control either through the pulse length or through
the delays between multiple pulses.  Such control based on the Floquet
potentials has been demonstrated in H$_2^+$~\cite{Esry}.  Control is also
possible, in principle, for processes having two or more pathways to the same
final state.  In this case, some mechanism to control the phase accumulated
along each pathway is needed.

Figure~\ref{fig:pot_curves_ads} shows the adiabatic potential curves for
different
laser frequencies and field strengths.
These potentials show field strengths corresponding to
3.5$\times$10$^{12}$~W/cm$^2$ (${\cal E}_0$=0.01~a.u.~\cite{Units}),
8.8$\times$10$^{13}$~W/cm$^2$ (${\cal E}_0$=0.05~a.u.), and
3.5$\times$10$^{14}$~W/cm$^2$ (${\cal E}_0$=0.1~a.u.).  The range of frequencies
represented, $\omega$=0.2--0.6~a.u., mean that the number of photons required
for ionization range from three to one, respectively.  In our adiabatic Floquet
potentials, then, ionization of H($1s$) proceeds either by tunneling through a
barrier or by passing over it.

In addition to barriers, the adiabatic potentials in
Fig.~\ref{fig:pot_curves_ads} also have wells.  The well most likely to play an
important role in ionization of H$(1s)$ forms the other half of the avoided
crossing that produces the barrier discussed above.   Its minimum lies between
$r$=1 and 2 a.u., depending on $\omega$, but is energetically accessible at
--0.5~a.u. only for $\omega$=0.4 and 0.6~a.u. in the figure.  When it is
energetically accessible, there is a chance that part of the initial wavepacket
will be trapped in the well --- the analog of molecular vibrational trapping.
Should this happen, it might be observable as a reduced ionization rate.  Or, in
a time-dependent measurement, it might be observable as ionization on different
time scales as the trapped portion of the wavepacket should take longer to
ionize.  Other analogs to molecular phenomena can be identified, some of which
will be described below.

The discussion of these adiabatic Floquet potentials would not be complete
without some comment on their range of applicability.  Even though the adiabatic
Floquet representation described here and implemented in Eq.~(\ref{RadEqns})
is exact, the potentials themselves form a useful qualitative picture only when
there are not too many of them.
Physically, this condition is clearly achieved if the adiabatic parameter $r$ is
indeed the ``slowest'' variable in the system.  Practically, so long as the
photon energy is not too much smaller than the ionization energy, then the
adiabatic picture will be useful.  For H($1s$), the $\omega$=0.2~a.u. case shown
in Figs.~\ref{fig:pot_curves_ads}(a-1)--(a-3) is about as small as is useful.
Decreasing $\omega$ further not only requires a larger range of $n$, but also a
larger range of $l$ since $l\geq |n|$ follows from the dipole selection rule.
The complexity of the picture can thus grow rapidly.

\subsubsection{AC Stark Shifts}

\begin{figure}
 \includegraphics[clip=true,scale=0.6]{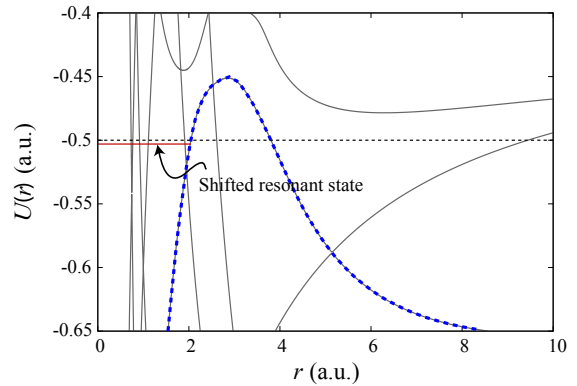}
\caption{(color online.) The diabatized potential curve (thick dashed blue line)
supporting the AC Stark-shifted ground state. Here, $\omega=0.2$ a.u.~and ${\cal E}_0=0.05$
a.u.}
\label{fig:pot_curve_ad_dia}
\end{figure}

The adiabatic Floquet potentials can also be used to make nontrivial
quantitative predictions without full-blown calculations.  One such example is
the AC Stark shift.  Strictly speaking, in a laser field, all bound states
become resonances --- a point made clear within the present representation.  The
shift in the position of that resonance from the field-free value is the
AC Stark shift.  Since our adiabatic potentials include the effect of the laser
field, we will estimate the ground state energy in our initial channel using a
simple WKB formalism (including the Langer correction~\cite{Nielsen1999}),
\begin{equation}
\int_{r_1}^{r_2}dr \sqrt{2\left[\varepsilon_{1s}-V(r)-\frac{1}{8 r^2}\right]}
=\frac{1}{2}\pi,
\label{WKB}
\end{equation}
which is
applicable only when the shifted state is still below the barrier shown in
Fig.~\ref{fig:pot_curves}.
Here, $r_1$ and $r_2$ are the classical turning points, and $\varepsilon_{1s}$
is the shifted ground state energy. Because the initial {\em adiabatic}
potential in Fig.~\ref{fig:pot_curves} is, in fact, quite complicated with very
sharp avoided crossings at $r$ values smaller than the barrier position, we used
a diabatized potential $V(r)$ in the calculation (see
Fig.~\ref{fig:pot_curve_ad_dia}).  This approximation can be physically
justified by the fact that these sharp crossings are most likely traversed
diabatically.

\begin{figure}[!htp]
 \includegraphics[clip=true,scale=0.33]{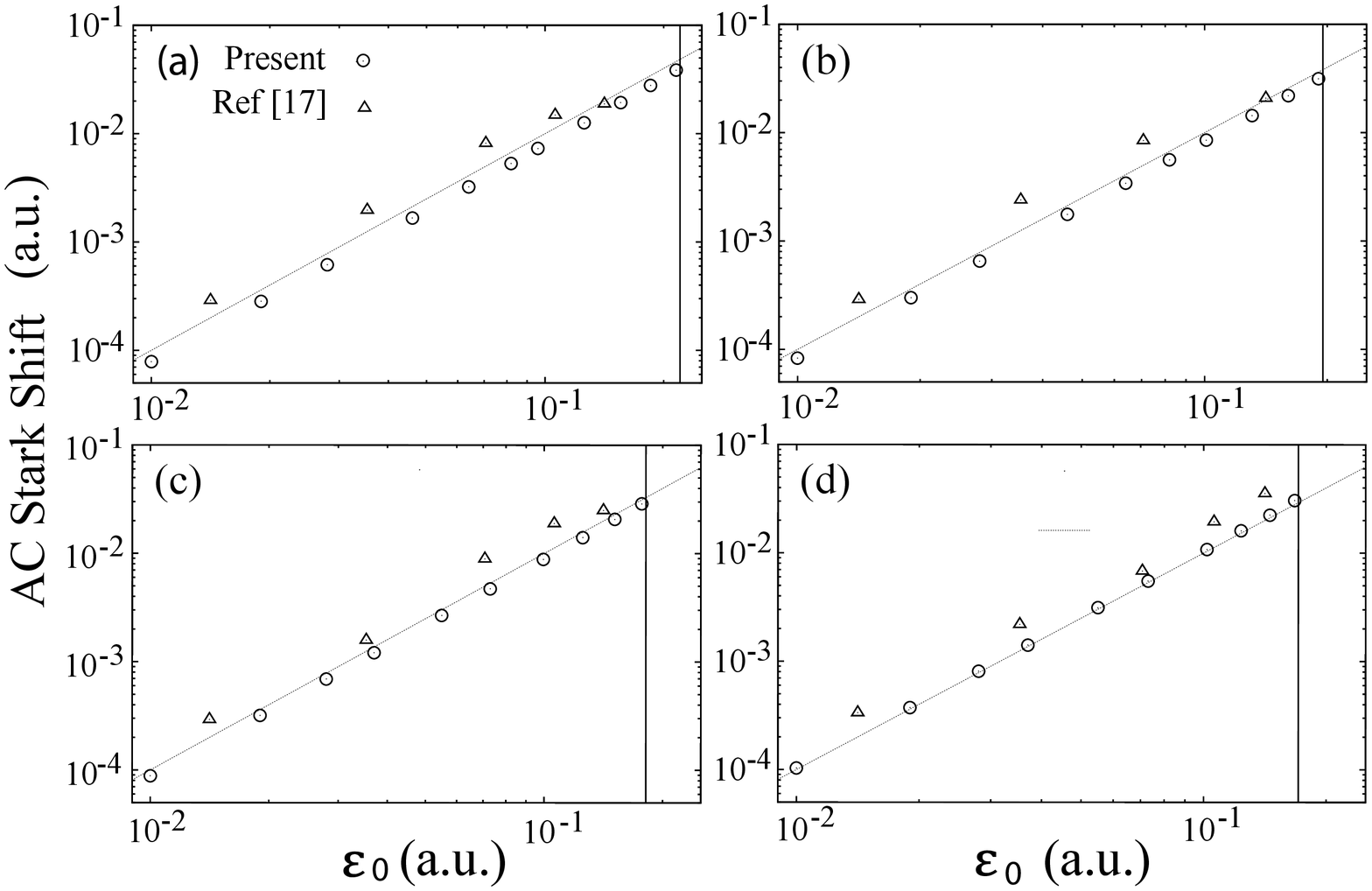}
\caption{The absolute value of ground state energy shifts of a hydrogen atom in
a linearly polarized
laser field as a function of field strength for (a) At $\omega=0.18\,$a.u.
(b) At $\omega=0.19\,$a.u. (c) At $\omega=0.20\,$a.u.~and (d) At $\omega=0.22\,$a.u.
Note the actual shifts are negative.
The dotted line shows ${\cal E}_0^2$ for comparison.
The vertical line indicates the field at which over-the-barrier ionization begins.}
\label{fig:shift_intensity}
\end{figure}
The ground energy shifts as a function of field strength for several $\omega$ are
shown in Fig.~\ref{fig:shift_intensity}.
For comparison, we also include the results calculated by the non-Hermitian
Floquet matrix method~\cite{Chu_groundhydrogen}, which predicts a larger shift
than ours in all cases.
Surprisingly, our results give nearly perfect quadratic behavior of the shifts with ${\cal
E}_0$ in the regime where the barrier is still present.  Quadratic behavior is
expected for the shift from second-order perturbation theory.
The differences between our results and the non-Hermitian Floquet
calculations, which should be quite accurate, are most likely due to our neglect of the
non-adiabatic couplings in Eq.~(\ref{NonAd}).

\subsubsection{Ionization Rates}
\begin{figure}[!htp]
 \includegraphics[clip=true,scale=0.6]{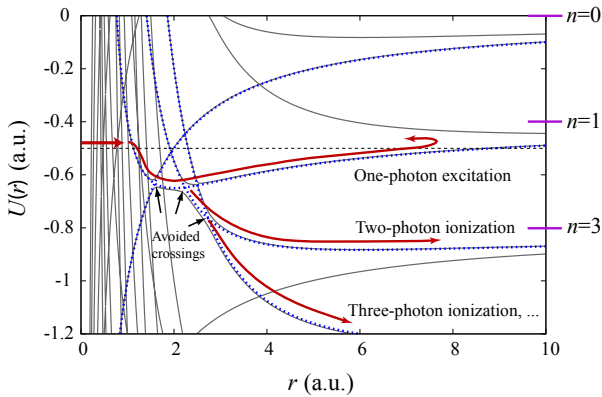}
\caption{(color online.) The pathways for a ground state wavepacket in a linearly polarized laser
field.
The horizontal arrow indicates the initial state at an energy of $-0.5$ a.u.~(dashed line).
Here $\omega=0.4$ a.u.~and ${\cal E}_0=0.05$ a.u.}\label{fig:path}
\end{figure}
Another quantity that can be estimated from our adiabatic potentials is the
ionization rate.
The ionization rates are calculated for the cases where the ground state energy
is higher than the top of the barrier shown in the
above adiabatic potential curves.
We use the Landau-Zener transition formula~\cite{Landau_Zener} for curve
crossings to obtain the transition probabilities.
The frequency
regions where our calculations of ionization rates could apply are restricted by
the prerequisites of the Landau-Zener formula:
the energy of the wave-packet is high above the curve crossing point and only
two-curve crossings are involved. Figure~\ref{fig:path}
shows the corresponding pathways for the present purpose. Since the first
crossing point (between diabatic channels $l=0,n=0$ and $l=1,n=1$) is lower than
the ground state
energy when $\omega > 0.25 $, $\omega$ needs to be larger than
$0.25\,$a.u when the system is initially in the ground state,
for the transition formula to be applied.

Strictly speaking, to get the ionization rates we need to include all the
adiabatic potentials and the couplings between them,
which would be tough numerically because of the sharp avoided crossings.
Fortunately, we can treat most of the sharp crossings as
purely diabatic. We also simplify the potential curves by seeking the most
important pathways relevant to the physical process we are
interested in. Here we select sequential one-photon processes, hence neglecting
simultaneous multi-photon absorptions(emissions).

Since Landau-Zener only provides the
the transmission probabilities,
the flux of electrons through the crossing is needed to calculate the ionization rates.
We determine it from the
classical argument that an electron with $l$=0 and an energy
of $-0.5$ a.u.~passes the crossing point twice
when it moves from one end of its trajectory to the other. The frequency of
the electron passing through the crossing point is thus $\Omega=1/\pi$\,a.u.
The ionization rate is then obtained as following:
The rate of ionization to the $n$-th Floquet channel is then
\begin{equation}
W_n=\Omega \left(\prod_{l=0}^{n-1}T_{l,l+1}\right)\,(1-T_{n,n+1}),
\label{eq:ionization}
\end{equation}
taking into accounr the absorption of each of the $n$ photons.
The Landau-Zener transition probability is
\begin{equation}
T_{l,l+1}=\left[ 1-
\exp\left(-\frac{2\pi}{\hbar}\frac{V_{l,l+1}^2}{F_{l,l+1}v_{l,l+1}}\right)
\right],
\label{eq:Zener}
\end{equation}
$V_{l,l+1}$ is the diabatic coupling,
$F_{l,l+1}$ is the classical ``force'',  and $v_{l,l+1}$ is the ``velocity''.
They are all evaluated at the crossing $r_{l,l+1}=\sqrt{(l+1)/\omega}$ and are given by
\begin{align}
V_{l,l+1}=\sqrt{\frac{\pi}{3}\frac{l+1}{\omega}}\langle Y_{l 0}|Y_{1
0}|Y_{l+1,0}\rangle {\cal E}_0,\nonumber\\
F_{l,l+1}=\frac{2\,\omega^{\frac{3}{2}}}{\sqrt{l+1}},\nonumber \\
v_{l,l+1}=\sqrt{2\left( \varepsilon_{1s}+\frac{l}{2}\omega+\sqrt{\frac{\omega}{l+1}}\right)}.
\label{eq:definition}
\end{align}
The ionization rates for each threshold shown in the adiabatic potential curves
for $\omega=0.6$~a.u.
in Fig.~\ref{fig:pot_curves_ads} are
calculated by Eq.~(\ref{eq:ionization}).
The outgoing kinetic energy of the ionized electron is
simply obtained by finding the difference between the initial energy
(ground-state energy of a field-free hydrogen atom) and the
outgoing threshold $-n\omega$.
\begin{figure}[!htp]
 \includegraphics[clip=true,scale=0.6]{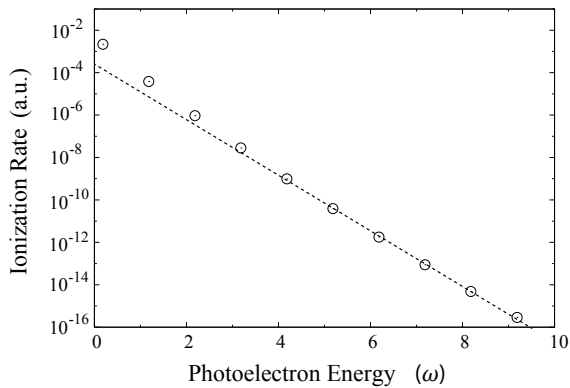}
\caption{The ionization rates as function of photoelectron energy at
$\omega=0.6$\,a.u., ${\cal E}_0=0.1$\,a.u.~for a linearly polarized field.
The dashed line indicates a pure exponential decrease for comparison.
}
\label{fig:transitions}
\end{figure}
\begin{table}
\caption{Comparison of ionization rates for H($1s$) at $\omega=0.6\,$a.u. in a linearly polarized field.}
\begin{ruledtabular}
\begin{tabular}{ccc}
 ${\cal E}_0$(\,a.u.) &$\Gamma/2$\footnote{From present calculation}(\,a.u.)
&$\Gamma/2$\footnote{From non-Hermitian Floquet
method~\cite{Chu_groundhydrogen}}(\,a.u.)\\
\hline
$\sqrt{2}\cdot 0.01$  & $0.3937\times 10^{-5}$ & $0.1253\times 10^{-3}$\\
$\sqrt{2}\cdot 0.025$  & $0.1414\times 10^{-3}$ & $0.7836\times 10^{-3}$\\
$\sqrt{2}\cdot 0.05$  & $0.5618\times 10^{-3}$ & $0.3144\times 10^{-2}$\\
$\sqrt{2}\cdot 0.075$  & $0.1250\times 10^{-2}$ & $0.7107\times 10^{-2}$\\
\end{tabular}
\end{ruledtabular}
\label{tab:width}
\end{table}

Figure~\ref{fig:transitions} shows the ionization rates for
multi-photon absorption up to $n=10$. Since in our picture
the laser field is purely monochromatic, the outgoing kinetic energies can only
have the discrete values $E_n=n\omega-\varepsilon_{1s}$,
which corresponds to the peaks of the above threshold ionization (ATI) spectrum.
It is seen from Fig.~\ref{fig:transitions}
that the rates drop nearly exponentially as $n$ increases, which is the same
feature seen in ATI experiments~\cite{ATI}.
We also converted our total ionization rates to the
half-width of the shifted ground-state energy for a rough comparison with the
data from~\cite{Chu_groundhydrogen}.
It can be seen in Table~\ref{tab:width} that our results 
are on the same order of magnitude as the results from
Ref.~\cite{Chu_groundhydrogen},
which is very good
agreement considering the simple scheme we used.

\subsection{Circularly Polarized Field}
Our approach can be generalized to circularly polarized laser fields.
In this case, we take $z$ axis of the electronic
coordinate to be perpendicular to the laser polarization,
such that $\boldsymbol{{\cal E}}={\cal E}_0[\cos(\omega t)\hat{x}+\sin(\omega
t)\hat{y}]$.
The matrix elements of ${\cal H}_{\rm
ad}(r)$ are now
 \begin{widetext}
\begin{align}
\langle\!\langle nlm |{\cal H}_{\rm ad}|n'l'm'\rangle\!\rangle&=\delta_{n
n^\prime}\delta_{l l^\prime}\delta_{m m^\prime}
\left[\frac{l(l+1)}{2r^2}-\frac{1}{r}-n\omega\right]+ \nonumber \\
&\Big[-\delta_{n,n'+1}\langle
Y_{l m}|Y_{1 1}|Y_{l' m'}\rangle
+\delta_{n,n'-1}\langle Y_{l m}|Y_{1, -1}|Y_{l'
m'}\rangle\Big]\sqrt{\frac{2\pi}{3}}r {\cal E}_0.
\label{Adia_Hamil_CP}
\end{align}
 \end{widetext}
Again the number of potentials curves we need to include can be simplified by selection rules.
As with the linearly polarized laser filed, only those diabatic
channels with both $n$ and $l$ even or
$n$ and $l$ odd are allowed.
For circularly polarized laser field, however, $m$ is now restricted by $m=n$.

\begin{figure}
 \includegraphics[clip=true,scale=0.6]{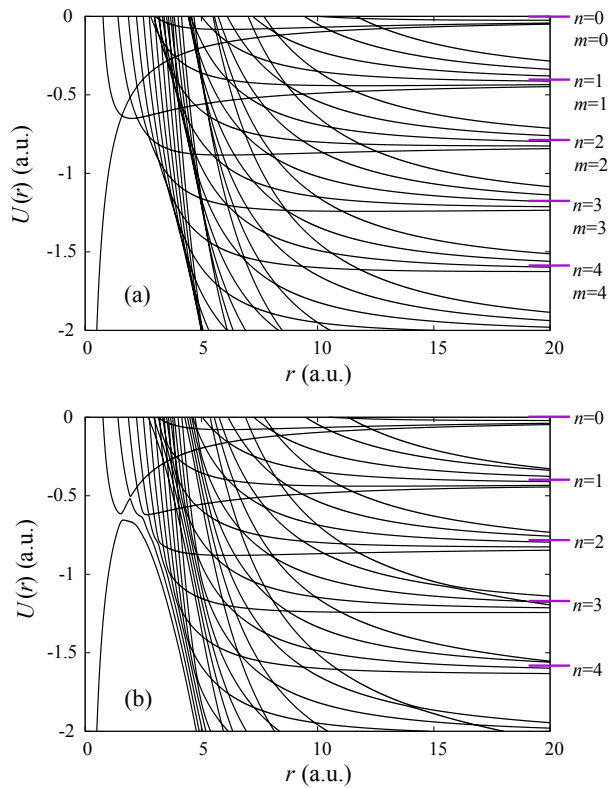}
\caption{(color online.) (a) The diabatic Floquet potentials for a hydrogen atom
in a circularly polarized field with $\omega=0.2$\,a.u.
The thick blue curve supports the initial state.
(b) The corresponding adiabatic Floquet potentials for ${\cal E}_0=0.05$\,a.u.
The Floquet blocks included here are $-11\leq n\leq 11$ and the angular momentum
states included are $0\leq l\leq 11$.
}
\label{fig:pot_circular}
\end{figure}

We show the diabatic and adiabatic potentials for a circularly polarized laser field
in Fig.~\ref{fig:pot_circular}.
There is a significant difference in these the potentials
from the
linearly polarized case:
at any value of $r$, there is always a lowest potential curve no matter how large $n$ becomes.
This is certainly not true in for linear polarization since increasing $n$ always adds a
lower energy potential.
The difference stems from the fact that in a
circularly polarized field $m=n$
which makes the angular momentum $l$ increase as $n$
increases. Since the diabatic potentials are quadratic
in $l$ but linear in $n$, they become increasingly repulsive for larger $n$. As illustrated in
Fig.~\ref{fig:pot_circular}(b), we observe that at large distances, the lowest
potentials become parallel to each other and diverge to $-\infty$ like $-r^2$.
The electron will therefore accelerate to infinity if it stays in one adiabatic channel. We interpret this unphysical 
situation by failure of the aiabatic approximation for large $r$ in the circularly polarized field.
We first note that in our picture, the change of the adiabatic potentials signifies the gain of electronic energy from the laser field 
in the region of large $r$ where the Coulomb potential is negligible. 
This energy gain can be understood by the following
simple classical argument: the way we calculate the adiabatic potential curves is equivalent to confining the 
electron to a fixed $r$ and finding its stationary motion.  In a circularly polarized field, the stationary motion is just that the
electron moves in a circle in constant speed with the electric force always perpendicular to the direction of the motion. 
The eletron energy for large $r$ is then given by $\omega^2 r^2/2$, in agreement with the leading order of $|U(r)|$.
The physical classical motion of the electron during ionization, however, is a spiral trajectory where the asymptotic energy of the electron 
does not change drastically.
Therefore, to correctly describe the motion of the electron in the adiabatic Floquet picture, 
a lot of adiabatic channels need to be coupled for large $r$, which makes pathway analysis impractical.

\begin{figure}
 \includegraphics[clip=true,scale=0.6]{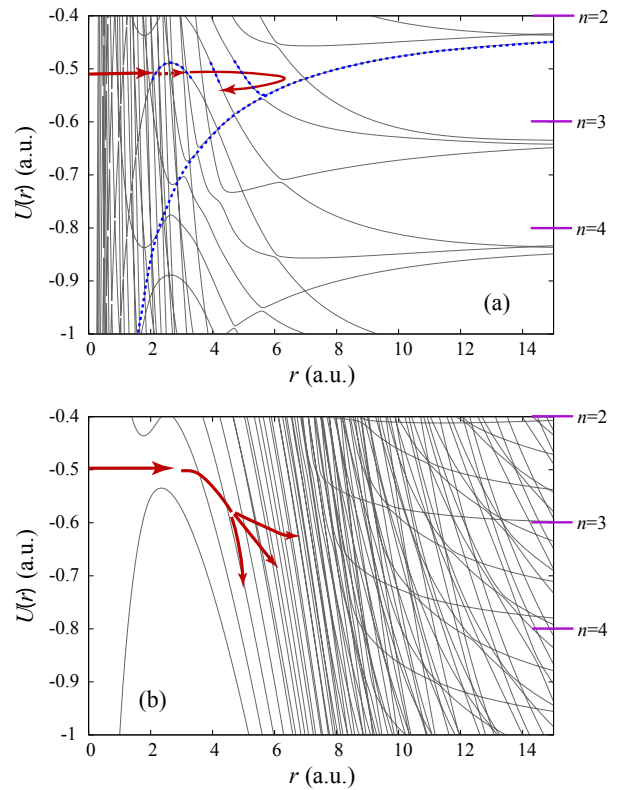}
\caption{(color online.)
Rescattering dynamics for (a) linearly and (b) circularly polarized light.
The red arrows indicate the ionization pathway most relevant to rescattering,
while the blue dashed curves trace the corresponding potential. In (a), $|n| \le 15$
Floquet blocks  were included with $l\le 5$; in (b), $|n| \le 21$ and $l \le 21$.  In both
figures, $\omega=0.2$ a.u.~and $\mathcal{E}_0 = 0.05$ a.u.
}
\label{fig:pot_path}
\end{figure}
\begin{figure}
 \includegraphics[clip=true,scale=0.6]{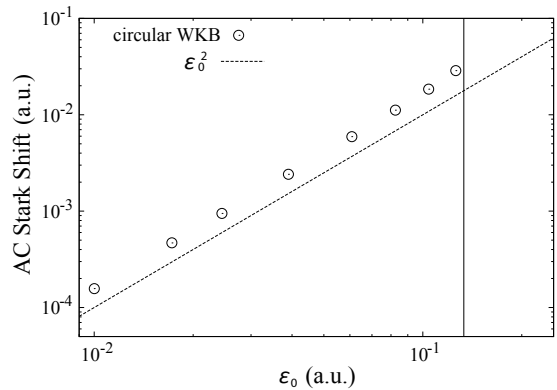}
\caption{(color online.) The absolute value of ground state energy shifts of
hydrogen atom in circularly polarized laser field
as function of field strength at $\omega=0.18$ a.u.. The dashed line shows ${\cal
E}_0^2$ as comparison of the behavior of the shifts,
the vertical line shows the field where above of barrier ionization occurs.}
\label{fig:shift_cir}
\end{figure}

In much the same way as for linear polarization,
we can calculate the AC Stark shift
of the H($1s$) ground state in a circularly polarized
field. Now though, the lowest adiabatic potential can be directly used without
diabatization. In Fig.~\ref{fig:shift_cir}, we show
the AC Stark shifts calculated by Eq.~(\ref{WKB}). Similar to the linearly
polarized case, the shifts follow an ${\cal E}_0^2$
scaling.
Comparing with Fig.~\ref{fig:shift_intensity}(a) for the linearly polarized case,
we see that our approximation predicts the AC Stark shifts for circular polarization
are larger than for linear polarization, even when compared at equal intensities,
i.e.~$\mathcal{E}_\text{lin} = \sqrt{2}\mathcal{E}_\text{circ}$.

When one-photon ionization is dominant, we can calculate the ionization rates in
a similar manner as discussed for linearly polarized fields in Sec. IIIC.
The potentials and transitions are identical to those but with different couplings.
The expression for the ionization
rates is then still given by Eqs.~(\ref{eq:Zener})
and (\ref{eq:definition}), with $V_{l,l+1}$ replaced by
\begin{equation}
V_{l,l+1}=\sqrt{\frac{2\pi}{3}\frac{l+1}{\omega}}\langle Y_{l+1,l+1}|Y_{1
1}|Y_{l l}\rangle{\cal E}_0.
\end{equation}
The first diabatic coupling $V_{0,1}$ has the same value as in a
linearly polarized field if the field ${\cal E}_0$
is scaled by $1/\sqrt{2}$, which correspond to the field with the same
intensity. 
Since the total ionization rate for linearly and circularly polarized fields in the perturbative limit are dominated by
the single-photon process, they should be related by a simple rescaling of the field strength.

Having potentials for both linearly and circularly polarized light, one interesting
question to consider is whether they can clearly describe the difference in
rescattering behavior between the polarizations.  Namely, electrons ionized in a linearly
polarized field can be driven to rescatter from their parent ion; electrons ionized by a
circularly polarized field do not rescatter.  To answer this question, we show in Fig.~8 the adiabatic
Floquet potentials for both polarizations.  For linear case in Fig.~8(a), the
electron tunnels through the barrier to ionize.  Some part of this ionizing wavepacket can find
itself on the $n=2$, $l=0$ curve after a few transitions.  This potential is attractive, and this
part of the the wavepacket gets reflected back towards the ion as expected.  For the circular case in
Fig.~8(b), however, once the electron passes over the barrier to ionize, it encounters only repulsive potentials
and thus never rescatters.

\section{\bf Summary}
In this paper we have developed the adiabatic Floquet picture to describe the
dynamics for hydrogen atom in monochromatic laser
field. although the analysis is trivially extended to any effective one-electron atom.
The radial distance of the electron
$r$ is adopted as adiabatic parameter, leading to a set of potential curves that include the
effect of the laser field.
The potentials are completely analogous to Born-Oppenheimer potentials can be interpreted in the
same way.  In particular, very simple approximations such as WKB and Landau-Zener can be applied
to make semi-quantitative predictions from these potentials that are non-perturbative in the field
strength.

Just how fruitful this picture is remains to be seen.  The diabatic curves have the distinct advantage of being
very easy to generate, and already allow the application of simple approximations.  While we did not explore it here,
one of the more intriguing possibilities is using these curves to estimate the times at which particular
photon transitions occur.  Few other approaches provide this information, especially in such a simple manner.
Having established the reliability of our picture in this paper, we hope that such timing information
can prove useful in controlling attosecond dynamics.

\appendix
\section{Gauge issues}
\label{App_Gauges}

While physical observables must be
gauge invariant (in the limit of no approximations), the adiabatic Floquet
potential curves are not.  The choice of gauge is usually made for computational
convenience and clarity in analysis.
There are, in principle, an infinite number of choices, but in practice
three main gauges are usually discussed:
(\textit{i}) the velocity gauge, which is the minimal-coupling Schr\"odinger
equation in the dipole approximation; (\textit{ii}) the length gauge, which was
the starting point for this paper; and (\textit{iii}) the so-called
``acceleration gauge'' (also known as
the Kramers-Henneberger frame \cite{kramers,henneb}).  As mentioned above, the
velocity gauge is not attractive for an adiabatic representation because not
all terms in the Hamiltonian involving the field are contained in ${\cal H}_{\rm
ad}$.  The length and acceleration gauges, however, are both good candidates for an
adiabatic picture since the laser field is completely included in ${\cal
H}_{\rm ad}$ (see also \cite{HMiyagi2010}).  Here we
will briefly discuss the adiabatic Floquet picture for atomic hydrogen in the
acceleration gauge.

The adiabatic Hamiltonian for hydrogen in the acceleration gauge is
\begin{equation}
 	{\cal H}_{\rm ad} = \frac{{L}^2}{2r^2} - \frac{1}{|{\bf
r}-{\mathbf\alpha}(t)|} - i\frac{\partial}{\partial t},
	\label{eq:appx01}
\end{equation}
where ${\mathbf\alpha(t)}$ is the classical trajectory of a free electron in an
oscillating field.  In the present case of a CW laser,
\begin{equation}
 	{\mathbf\alpha}(t) = \hat{z}\alpha_0 \cos(\omega t)
	\label{eq:appx02}
\end{equation}
with $\alpha_0 = {\cal E}_0/\omega^2$, assuming $\alpha(0)=\dot\alpha(0)=0$.

One advantage of the acceleration gauge is that the
effect of the electric field vanishes in the limit $r\rightarrow\infty$.
Understandably, then, the usual field-free solutions are recovered,
whereas in the length gauge the laser-atom interaction diverges linearly in $r$.
 A disadvantage of the acceleration gauge  is the difficulty presented by the
oscillating Coulomb singularity.

Applying the same adiabatic Floquet analysis as in Sec.~\ref{Theory},
we obtain the following matrix elements for the adiabatic
Hamiltonian:
\begin{widetext}
\begin{equation}
\langle\!\langle nkm |{\cal H}_{\rm ad}|n'k'm'\rangle\!\rangle
 = \delta_{n n\prime}\!\int_0^\pi\! \sin\theta\,u_{k^\prime}^*(\theta)\left(-n
\omega +\frac{{L}^2}{2 r^2}
 \right)u_k(\theta)d\theta -
  \frac{1}{T}\int_0^\pi\! \int_0^T \!\frac{ \sin\theta\, u_{k^\prime}^*(\theta)
e^{-i(n-n^\prime)\omega t} u_k(\theta)}
{\sqrt{r^2-2r\alpha_0\cos(\omega t)\cos(\theta)
  + \alpha_0^2\cos(\omega t)}}dt d\theta,
\end{equation}
\end{widetext}
where $n$ and $n^\prime$ are again the Floquet indices, and $u_k(\theta)$
replace the spherical harmonics as the basis functions for the $\theta$
dependence.  In our calculations, we took $u_k(\theta)$ to be b-splines~\cite{EsryThesis}.

\begin{figure}
  	\includegraphics[scale=0.9]{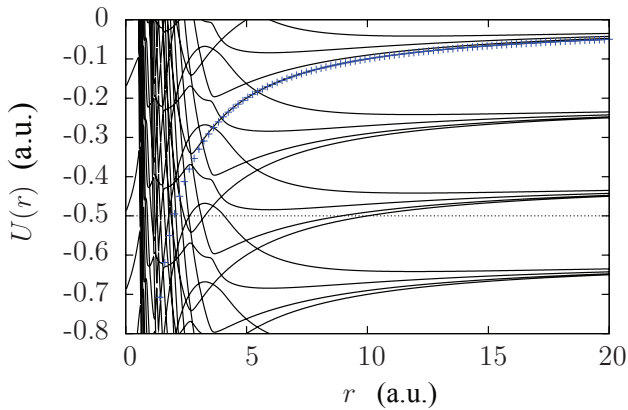}
	\caption{The adiabatic Floquet potentials for atomic hydrogen in the
acceleration gauge.  The crosses indicate the initial curve the system
				starts on when the electric field is zero
($\alpha_0 = 0$).
				19 Floquet blocks were used in this calculation
(centered around $n=0$).
	}
	\label{fig:accel_pot}
\end{figure}

In Fig.~\ref{fig:accel_pot}, we plot the adiabatic Floquet potentials in the
acceleration gauge, using the same parameters as in
Fig.~\ref{fig:pot_curves}(b), namely $\omega$=0.2~a.u. and ${\cal
E}_0$=0.05~a.u.

At that electric field strength and frequency, $\alpha_0 = 1.25$, and in the
figure the complicated
nature of the curves is quite clear for $r<1.25$.  While in the acceleration
gauge, the channels decouple at large $r$, they are strongly coupled
at small $r$.
Further, unlike the length gauge, the familiar dipole selection rules are no
longer valid, and in fact
all of field-free angular momentum states are coupled together near
$r=\alpha_0$.  This makes tracing a path through the potential curves
very difficult, even locating the curve the system initially sits on is a
challenge. The crosses in Fig.~\ref{fig:accel_pot} trace out the initial state
when $\alpha_0 = 0$.
In the underlying mess of curves, one can see the semblance of the avoided
crossings mentioned in Fig.~\ref{fig:pot_curves}(b), but the hopes of
finding a path that passes though these crossings is daunting.
While the thresholds are easy to determine in the acceleration gauge, the steep
trade-off is in the loss of physical intuition in the region close
to $r=\alpha_0$.

\begin{acknowledgments}
This work is supported by the Chemical Sciences, Geoscience, and Biosciences
Division,
Office for Basic Energy Sciences,Office of Science, U.S. Department of Energy.
Y. W. also acknowledge the support from the National Science Foundation under Grant No. PHY0970114.
\end{acknowledgments}


\begin{thebibliography}{99}

\bibitem{Corkum1993} P. B. Corkum, Phys. Rev. Lett. {\bf 71}, 1994 (1993).

\bibitem{KSK1992} J.~L.~Krause, K.~J.~Schafer, and K.~.C.~Kulander, Phys. Rev A, {\bf 45} 4998 (1992).

\bibitem{Lewenstein1994} M. Lewenstein, Ph. Balcou, M. Yu. Ivanov, Anne
L’Huillier, and P. B. Corkum, Phys. Rev. A {\bf 49}, 2117 (1994).

\bibitem{Esry} B. D. Esry, A.M. Sayler, P. Q. Wang, K. D. Carnes, and I.
Ben-Itzhak, Phys. Rev. Lett. {\bf 97}, 13003 (2006).

\bibitem{AStaud:PRL2007} A.~Staudte, {\it et al.} Phys. Rev. Lett. {\bf 98}, 073003 (2007).

\bibitem{BEsry:PRA2010} B.~D.~Esry and I.~Ben-Itzhak, Phys. Rev. A {\bf 82}, 043409 (2010).

\bibitem {JHern:BAP2003} J.~V.~Hern\'andez and B.~D.~Esry, Bull. Am. Phys. Soc. {\bf 48}, 123 (2003).

\bibitem{Chu_review} S.-I. Chu and D. A. Telnov, Phys. Rep. {\bf 390}, 1 (2004).

\cite{JShir:PR1965,RJens:PR1991,ABuch:PR2002,HMaed:PRL2006}

\bibitem{JShir:PR1965} J.~Shirley, Phys.~Rev.~{\bf 138}, B979, (1965).

\bibitem{RJens:PR1991} R.~V.~Jensen, S.~M.~Susskind, and M.~M.~Sanders, Phys.~Rep.~{\bf 201}, 1 (1991).

\bibitem{ABuch:PR2002} A.~Buchleitner, D.~Delande, and J.~Zakrewski, Phys.~Rep.~{\bf 368}, 409 (2002).

\bibitem{HMaed:PRL2006} H.~Maeda, J.~H.~Gurian, D.~V.~L.~Norum, and T.~F.~Gallagher, Phys. Rev. Lett.~{\bf 96}, 073002 (2006).

\bibitem{HMiyagi2010} H. Miyagi and K. Someda, Phys. Rev A {\bf 82}, 013402
(2010).

\bibitem{Intro1} M. Gavrila (Ed.), Atoms in Intense Laser fields, Academic
Press, New York, 1992.

\bibitem{Units} We define 1\,a.u. of electric field to correspond to $3.51\times
10^{16}~\mathrm{W/cm}^2$ of intensity for linearly polarized laser field.

\bibitem{Nielsen1999} E. Nielsen, and J. H. Macek, Phys. Rev. Lett. {\bf 83},
1566 (1999).

\bibitem{Chu_groundhydrogen} Shih-I Chu and J. Cooper, Phys. Rev. A {\bf 32},
2769 (1985). They define $1.0$\,a.u. of rms field strength
corresponds to rms intensity of $7.0\times 10^{16} \mathrm{W/cm}^2$.

\bibitem{Landau_Zener} C. Zener, Proc. Roy. Soc. London {\bf 137}, 696 (1932).

\bibitem{ATI} R.R. Freeman, P.H. Buksbaum, W.E. Cooke, G. Gibson, T.J. McIlrath,
L.D. van Woerkom, Adv. At. Mol. Opt.Phys. Suppl. {\bf 1}, 43 (1992).

\bibitem{kramers} H. A. Kramers {\it Collected Scientific Papers} North Holland,
Amsterdam (1956).

\bibitem{henneb}  W. C. Henneberger, Phys. Rev. Lett. {\bf 21}, 838 (1968).

\bibitem{EsryThesis} B. D. Esry, {\it Many-body effects in Bose-Einstein condensates of dilute atomic gases}, Ph.D. Thesis (1997), App. C.

\end{thebibliography}
\end{document}